\documentclass[12pt,a4paper]{article}
\usepackage{graphicx}
\usepackage{amsmath}
\usepackage{graphicx}
\usepackage{amsfonts}
\usepackage{float}

\begin{document}

\author{S.Brazovskii$^{a,\ast}$, and N.Kirova$^{b}$ \\
$^{a}$LPTMS, UMR 8626, CNRS \& Universit\'{e} Paris-Saclay, France\\
$^{b}$LPS, UMR 8502, CNRS \& Universit\'{e} Paris-Saclay, France\\
 $^{\ast}$e-mail: brazov@lptms.u-psud.fr}

\title{Multi-fluid hydrodynamics in charge density waves with collective, electronic, and solitonic densities and currents.}
\date{\textit{Contribution for the special volume in honor of I.M. Khalatnikov 100th birthday. JETP, 129, 659 (2019)}}
\maketitle

\begin{center}
\textbf{Abstract}
\end{center}

We present a general scheme to approach the space - time evolution of deformations, currents, and the electric field in charge density waves related to appearance of intrinsic topological defects: dislocations, their loops or pairs, and solitons.
We derive general equations for the multi-fluid hydrodynamics taking into account the collective mode, electric field, normal electrons, and the intrinsic defects.
These equations may allow to study the transformation of injected carriers from normal electrons to new periods of the charge density wave, the collective motion in constrained geometry, and the plastic states and flows. As an application, we present analytical and numerical solutions for distributions of fields around an isolated dislocation line in the regime of nonlinear screening by the gas of phase solitons

\section{Introduction: CDWs and their intrinsic defects.}

\subsection{Theoretical and experimental motivations.}

Charge density waves (CDWs) are ubiquitous in quasi one-dimensional (1D) electronic systems \cite{Gruner:88,Gorkov+Gruner,Monceau:12}.
The CDW is a superstructure $\propto A\exp (\vec{q}_{0}\vec{r}+\varphi )$ within the host crystal produced by spontaneous modulations of the electronic charge and atomic
displacements. The CDW wave number $\vec{q}_{0}$ must be close to Fermi diameter $2k_{F}$ of the parent metal thus opening a gap  $\Delta \propto|A|$ in the spectrum of electrons. We shall always keep in mind the most interesting and common case of incommensurate CDW which period $\pi /k_{F}$ is not a rational number of the atomic one when the periodicity of the underlying lattice does not play a role. Such a CDW possesses a manifold of degenerate states which energies do not depend on a sign of the amplitude $A$ and on an arbitrary shift of the phase $\varphi $.
The translational degeneracy of the CDW ground state leads to the phenomenon of Fr\"{o}lich conductivity showing up in a giant dielectric susceptibility, in a collective conductance by virtue of sliding and in many related nonlinear and nonstationary effects \cite{Gruner:88,Gorkov+Gruner,Monceau:12}.

The degeneracy allows also for configurations connecting equivalent while different states across a disturbed area.
These configurations are typically topologically protected or at least are protected by the charge or spin conservation laws.
They are known commonly under a generic name of "topological defects" which includes here extended objects like planes of domain walls as arrays of solitons\cite{BK:84,SB:07}, lines or loops of dislocations \cite{Feinberg:88,SB+SM-disl:91,Hayashi:96} as phase vortices, and local objects like phase and amplitude solitons \cite{BK:84,SB:90,SB-Landau100:09}.
As well as conventional crystals, the CDWs are subjected to elastic and plastic deformations: the elastic ones follow external perturbations almost instantaneously while the reaction of plastic ones is limited by mobility and relaxation of intrinsic defects \cite{Andreev:92}. The intrinsic defects are genealogically related to electronic quasi-particles due to a strong interaction of CDW deformations with normal electrons which leads to their fast selftrapping in the course of conversion to various kinds of topologically nontrivial objects. Formation of these objects requires for an energy (per one electron or hole involved) which is below the minimal energy $\Delta $ for the band electron or the hole.
Hence the progressive energy release goes through a sequence of events of conversion of bare electrons \cite{SB:90}: electrons' self-trapping into amplitude solitons, fusion of pairs of electrons and/or amplitude solitons into phase solitons, aggregation (see \cite{Karpov:19} and refs. therein) of phase solitons into dislocation loops or their absorption by advancing dislocation lines.


In many theoretical and experimental observations the local defects, amplitude and phase solitons, compete with electrons as normal carriers of the spin and the charge correspondingly.
The phase solitons \cite{Horovitz:77,Artemenko:95,Artemenko-PRL:95} viewed as interstitials/vacancies of the CDW considered as a crystal of electronic
pairs, provide a material for the climb of dislocations required for their expansion.
The dislocations participate in depinning, in generation of coherent signals by a sequence of phase-slip events \cite{Gorkov:90,Ong:84}, in structures at contacts and junctions
\cite{SB+SM-cnt:91,SB+SM-cnv:92,NK+SB:05}.
Generation of intrinsic defects via sliding over the host lattice imperfections gives a clue \cite{Larkin:95,SB+TN:04,Miller:13} to observed nonlinear current-voltage characteristics.
Transverse proliferation of dislocations across the sample \cite{Ong:84} gives rise to phase slips \cite{Gorkov:90,Saint:85,Artemenko:86,Krive:87,Nad:89,Kovalev:96} which are necessary to bring about the CDW sliding along the sample.
Properties of all intrinsic defects at low temperature $T$ are strongly affected by Coulomb forces and depend on screening facilities of free carriers: electrons and solitons.

Experimentally, the presence of amplitude solitons could have been inferred from the spin susceptibility \cite{Johnson:83}, intrinsic tunneling \cite{Latyshev:05,Latyshev:06}, and via direct visualized by the STM \cite{Brun-soliton,yeom-soliton}. Phase solitons have been identified \cite{Zhilinskii:83,Nad:85,Matsuura:15}, via their low activation energy seen in the on-chain transport while the higher, probably from gaped electrons or holes, energy is responsible for the transverse transport and optics \cite{ZZ+VM:06} in which the solitons cannot participate. The presence of dislocations have been inferred from space-resolved X-ray studies of the sliding state (\cite{Lemay:98,Rideau:01} and refs. therein), from coherent X-ray micro-diffraction \cite{LeBolloch} also under sliding, and from reconstruction in mesa-junctions \cite{Latyshev:06}.

Fig.1 quoted from \cite{Brun-soliton} shows the STM image of the CDW containing one embedded defect seen here as the $-2\pi$ soliton: the
compression by one period along the defected chain is equivalent to a pair of dislocations embracing the chain.
Fig.2 quoted from \cite{Rideau:01} shows the profile of the CDW strain, the phase gradient $\partial_{x}\varphi$, appearing in the course of conversion among normal and collective currents near junctions and at an obstacle (a cross-section with a sharply increased pinning force).
Near the junctions at $x\approx\pm2$ and at the accidental obstacle at $x\approx0$, the intensive phase slips creation processes might take place by means of transverse flows of dislocations absorbing/emitting normal carriers and solitons. The dashed lines show a successful fitting based upon the theory of \cite{BK:99} which can also be derived as a simple limit of equations presented below in this article.

Describing the coexistence of electrons and defects in static equilibrium, under strains and in the current carrying state requires for a general nonlinear hydrodynamics for two fields - the phase and the electric potential and for two fluids of electrons and defects. Of special importance is to trace how the injected at the junction electrons finally evolve into deformation and collective motion of CDW in the bulk \cite{BK:99}. This article suggests a contribution to this request.

\subsection{Dislocations and solitons in Charge Density Waves. \label{CDW-Xl}}

With respect to deformations, the CDW is a kind of an elastic uniaxial crystal, which is characterized by distortions of the CDW phase $\varphi(\vec{r},t)$.
Reduction of standard definitions for elasticity of conventional crystals to the particular case of CDWs can be summarized as follows:

\medskip
\begin{tabular}[l]{ll}
	atomic density $\rho_{atomic}$ & modulation $A\cos(Qx+\varphi)$ \\
	displacements $\vec{u}/a\qquad$ & $-\vec{\nu}\varphi/2\pi~,~\vec{\nu}=(1,0,0) $ \\
	compression $\ \ -\nabla\vec{u}\qquad$ & charge $\partial_{x}\varphi/\pi$ \\
	velocity$\ \ \ \ \qquad\partial_{t}\vec{u}$ & current $-\vec{\nu}\partial
	_{t}\varphi/\pi$ \\
	vacancies or addatoms & $\mp2\pi$ solitons \\
	dislocations & phase vortices
\end{tabular}
\medskip

Already at the level of small deformations, the CDW shows specific properties coming from long range Coulomb forces.
The conventional energy of an anisotropic elastic media acquires a nonanalytic, usually dominating, contribution of an anomalous elasticity \cite{SB:90,SB+SM-disl:91,SB+SM-sol:91}. This big non-local contribution appears also in isotropic (typically 2D) Wigner crystals and is already well known for vortex lattices in superconductors, but in quasi-1D CDWs this energy cost may not be avoidable. For Wigner crystals and vortices with their 2D displacements $\vec{u}$, the long range energy can be eliminated prohibiting the "charged" configurations via a constraint $\vec{\nabla}\vec{u}=0$ which still leaves sufficient degrees of freedom with$\vec{\nabla}\times\vec{u}\neq0$.
In CDWs, where deformations in chains direction $\vec\nu=(1,0,0)$, with $\partial_{x}\varphi\neq0$ which is equivalent to $\vec{\nabla}\vec{u}\neq0$, are frequently the most necessary ones, paying the Coulomb energy cannot be avoided.

Even more interesting and demanding are peculiarities related to topologically nontrivial configurations which determine spectacular nonlinear properties of CDWs. As a system described by the complex order parameter $A\exp(i\varphi)$, the CDW is expected to maintain commonly known nonlinear objects: phase slips as space -time vortices already in D=1, point (in D=2) or line (in D=3) space vortices, bound vortex pairs (in D=2) or vortex rings (in D=3).
But, for the CDW being actually a crystal (of electronic pairs), the vortex becomes an edge dislocation \cite{Feinberg:88,SB+SM-disl:91} which is subjected to additional
constraint of the matter conservation.

A conventional motion of dislocations is allowed only as a matter conserving glide along the chains, in the direction of the Burgers vector which quantum here is $\vec{b}=2\pi\vec{\nu}$.
The transversal motion, the non conserving climb, is prohibited whatever is the driving force coming from the local stress - in a strong difference with respect to conventional vortices.
The climb becomes allowed, and even necessary, if the amorphous material is supplied to be condensed at the dislocation core.
In conventional crystals this is a very rare situation involving oversaturated gases of vacancies or addatoms.
In CDWs, the climb allowed by the condensation or liberation of normal carriers is the ultimate mechanism of conversion between normal and collective currents which is one
of the most important experimentally accessible phenomena.

Beyond vortices-dislocations as topologically nontrivial objects appearing under stresses or out of equilibrium, there are also their bound states with compensating topological charges: pairs (in D=2) or rings (in D=3) of vortices.
Lacking the topological protection, these objects nevertheless do not annihilate like in the conventional $XY$ model and the complex-field theory but are stabilized by the matter (here the number of condensed electrons) conservation law.
Having concentrated the vacuum charge of $2Ne$ where $N$ is the number of the encircled chains, for the minimal $N=1$ these are the charge $\pm2e$ objects in a form of $\pm2\pi$ solitons. In a discrete view of the quasi-1D system, the CDW at the defected chain gains or looses one period with respect to surrounding chains as it is visualized experimentally in Fig.1.
These phase solitons become the spinless $2e$- charged quasi-particles, which seem to be prone for a role of charge carriers.

Indeed, the energy of these solitons $E_{2\pi}\sim T_{c}$ comes from the interchain coupling \cite{Horovitz:77} measured by the observed transition temperature $T_{c}$ of the long range ordering which, by definition of the quasi one-dimensionality, is small with respect to the gap $\Delta$.
At first sight, this expectation correlates with a common observations \cite{Zhilinskii:83,Nad:85,ZZ+VM:06} of very low activation energies for the longitudinal conductivity with respect to high energies $\sim\Delta$ for the transverse one where the solitons are not allowed to contribute.
Nevertheless, a closer theoretical inspection recovers some paradoxes. While the local charge is indeed $2e$ generating the corresponding electric field, the long-range dipole perturbations exerted upon the surrounding chains are integrated into a totally compensating counter-charge \cite{SB+SM-sol:91} (the "momentum conservation law" for phase rotations). Then in average over any sample cross-section the soliton, and a dislocation loop more generally, cannot either carry the current or be driven by the homogeneous electric field. This paradox comes from an ill-defined partition among the local charge, its long-range tails and the overall collective mode.

The above mentioned issues call for a careful theoretical treatment for the ensemble of dislocations and their loops allowing for extracting their
averaged and integrated characteristics. The goal is to derive an effective hydrodynamic equations for multiple fields (distortions and potentials) and fluids (defects and normal carriers).
\section{Kinematics and averaging.}
\subsection{Local relations.}
We shall consider the kinematics at presence of moving dislocation lines or loops by adopting to CDWs \cite{SB+SM-disl:91,SB+SM-cnv:92} the prescriptions \cite{Kosevich,LLv7} of the continuous theory of dislocations.
At presence of vortices the local deformations and velocities $\omega_{j}$, $j=x,y,z,t$, are not derivatives of the same phase $\varphi$:
$\vec{\omega}\neq\vec{\nabla}\varphi$ if dislocations are present and moreover $\omega_{t}\neq\partial_{t}\varphi$ if they also move:

\begin{equation}
{\partial_{x}\varphi}\rightarrow\omega_{x}~,~
{\partial_{y}\varphi}\rightarrow\omega_{y}\, , \,{\partial_{z}\varphi}\rightarrow\omega _{z}\, , \,{\partial_{t}\varphi}\rightarrow\omega_{t}
\label{omega}
\end{equation}

The space- and space-time vorticities of the phase are characterized by uniquely defined vectors: the density $\vec{\tau}$ of dislocation lines and the flow $\vec{I}$ in the course of their motion or formation:
\begin{equation}
\vec{\tau}=\frac{1}{2\pi}[\vec{\nabla}\times\vec{\omega}]~,~\vec{I}=\frac {1}{2\pi}(\vec{\nabla}\omega_{t}-\partial_{t}\vec{\omega})~,~
\left[ \vec{\nabla}\times\vec{I}\right] +\partial_{t}\vec{\tau}=0
\label{tau+I}
\end{equation}
E.g. $\vec{I}=[\vec{\mathrm{v}}\vec{\tau}]$ for an element of the dislocation line moving with a velocity $\mathrm{\vec{v}}(t)$. Since $\vec{\nabla}\vec{\tau}=0$, $\vec{\tau}$ can be expressed via another function $\vec{P}$, the "dislocation moment density", as
\begin{equation}
\vec{\tau}=-[\vec{\nabla}\times\vec{P}]\ \mathrm{hence}\ \vec{\nabla}\times\vec{\omega}+{2\pi}\vec{\nabla}\times\vec{P}=0~
\mathrm{then}~\vec{\omega}+2\pi\vec{P}=\vec{\nabla}\varphi
\label{tau-p}
\end{equation}
Here we have introduced an arbitrary gauge function as the phase $\varphi$ because in absence of dislocations $\vec{\tau}\Rightarrow0$, $\vec{P}\Rightarrow0$, $\vec{\omega}\Rightarrow\vec{\nabla}\varphi$.
The CDW phase $\varphi$ can be restored as a single-valued, but not uniquely defined function.
The flow can be written from Eq.(\ref{tau+I}) as
\begin{equation*}
\vec{I}=\partial_{t}\vec{P}+\frac{1}{2\pi}\vec{\nabla}\left( \omega_{t}-\partial_{t}\varphi\right)
\end{equation*}
While $\vec{\omega}$ and $\vec{I}$ are uniquely defined functions of state, $\varphi$ is discontinuous and $\vec{P}$ and $\vec{\nabla}\varphi$ are singular at some surfaces $S$ based on the dislocation lines, see Fig.3. These surfaces may be chosen arbitrarily at a certain moment of time, and there is also a freedom in their subsequent time evolution. We shall use a freedom to fix the time dependent gauge of $\vec{P}$ in such a way that
\begin{equation}
\omega_{t}\equiv\partial_{t}\varphi~\mathrm{then}~\ \vec{I}=\partial_{t}\vec{P}~\mathrm{and}~ \vec{\nabla}\omega_{t}=\partial_{t}\left( \vec{\omega}+2\pi\vec{P}\right) = \partial_{t}\vec{\nabla}\varphi\;
\label{phi-t}
\end{equation}
This choice is especially convenient for CDWs where the dynamics is usually nonlinear in the velocity $\omega_{t}$. By definition, $\omega_{t},\vec{\omega},\vec{I}$ are singular on a dislocation line only; now the same holds also for $\partial_{t}\vec{P}$, hence the surface $S$ support for $\vec{P}$ is arbitrary only at some initial time.
Afterwards, $\vec{P}$ evolves only along the surface passed by dislocation line, that is following the trace of
physical singularities as shown in Fig.3.

\subsection{Averaged values.}

Consider now the media filled with small, while possibly distributed in sizes, dislocation loops (or pairs in 2D).
Primarily we keep in mind minimal loops encircling just one chain which are the $2\pi$ solitons, playing the role of minimal energy charge excitations in quasi-1D CDWs.
Each $2\pi$ increment within a bundle of $N$ chains encircled by a dislocation line adds  $N$ CDW periods, requiring to absorb $2N$ electrons to the CDW ground state.
The concentration of added periods, which we shall call the local density of defects $n_{d}$, is given by the volume density of dislocation line areas
projected to the $(y,z)$ plane $\perp\vec{\nu}$.
Our goal will be obtaining equations averaged over the microscopic density of such dislocation loops in terms of their average density $n_{d}$, the current (coming only from the
glide, hence directed along the chains) $\vec{j}_{d}=j_{d}\vec{\nu}$, and the production rate $dn_{d}/dt$.

We shall consider only edge dislocations lying in the $(y,z)$ plane, so that  $\vec{\tau}\bot\vec{b}||\nu$.
Only those are imperative for major properties of sliding CDWs while mixed (screw in addition to edge) components appear only as fluctuations or in the course of the glide when various segments of lines may move with different velocities meeting some obstacles. It seems that our results are valid at presence of these fluctuations also, provided that they are zero in average.
The two types of motion of dislocation lines, the glide and the climb, are illustrated in Fig.4.
The glide is a conservative (with respect to the number of encircled chains) motion - $dn_{d}/dt=0$, which is allowed only along $\vec{\nu}$: the current of defects can be only
one-dimensional $\vec{j}_{d}=j_{d}\vec{\nu}$ as describing their glide.
In this case $\vec{I}$, and with our choice of the gauge also $\partial_{t}\vec{P}$,  are orthogonal to the surface formed by the dislocation line during its motion in $x$ direction:
 $\partial_{t}\vec{P}=\vec{I}_{glide}=0$,  $\partial_{t}P_{y},\partial_{t}P_{z})=(0,I_{y},I_{z})$.
In case of the climb, the dislocation loop grows: the dislocation line moves transversely to $\vec{\nu}$  changing the number of encircled chains.
Then $\vec{I}_{climb}=\vec{\nu}I_{x}$ where $I_{x}=\partial_{t}P_{x}$ yields the increase of $n_{d}$ while there is no current of defects $j_{d}=0$.

For a plane dislocation line or loop (when $\tau_{x}=0$) we can write
\begin{equation*}
\vec{\tau}=\left(0,-\partial_{z}\tilde{P}_{x},\partial_{y}\tilde{P}_{x}\right) =[\vec{\nu}\vec{\nabla}]\tilde{P}_{x}
\end{equation*}
Considering the new function $\tilde{P}_{x}$ as a magnitude of the $x$- directed vector $\vec{\nu}\tilde{P}_{x}$, we find it to be just another possible gauge of $\vec{P}$.
In this case the discontinuity in $\varphi$ is distributed over the plane segment enclosed by the dislocation line and moves together with it, thus producing nonphysical singularities in $\partial_{t}\varphi$ which cannot exist in $\omega_{t}$.
The disadvantage of working with $\tilde{P}_{x}$ is that now $\omega_{t}\neq\partial_{t}\varphi$, but the advantage is that $\tilde{P}_{x}$ is a delta - function distributed over the plane area enclosed by the dislocation line, thus in average $\left\langle \tilde{P}_{x}\right\rangle =-n_{d}$ where $n_{d}$ is the density of defects - the total area of dislocation lines per unit volume. Finally we obtain a relation between the average density of dislocation lines and the related defects which is gauge independent:
\begin{equation}
\langle\vec{\tau}\rangle=-[\vec{\nu}\vec{\nabla}]n_{d}~,~
\langle\lbrack \vec{\nu}\vec{\tau}]\rangle=-\vec{\nabla}_{\bot}n_{d}
\label{<tau>}
\end{equation}

The conservation law for $n_{d}$ and $j_{d}$ reads $dn_{d}/dt=\partial_{t}n_{d}+\partial_{x}j_{d}$.
For the average of the dislocation flow $\left\langle \vec{I}\right\rangle $ we find from Eqs.(\ref{tau+I})
\begin{equation*}
\left[ \vec{\nabla}\times\left( \left\langle \vec{I}\right\rangle +\vec{\nu }\partial_{t}n_{d}\right) \right] =0~\mathrm{hence}~
\left\langle \vec {I}\right\rangle +\vec{\nu}\partial_{t}n_{d}=\vec{\nabla}g
\end{equation*}
To determine the function $g$ we notice that
\begin{equation*}
\partial_{t}n_{d}-\partial_{x}g=-\left\langle I_{x}\right\rangle
=-\left\langle \partial_{t}P_{x}\right\rangle =\partial_{t}n_{d}+\partial
_{x}j_{d}
\end{equation*}
Hence $\partial_{x}(g+j_{d})=0$, then $g=-j_{d}+\tilde{g}(\vec{r}_{\bot},t)$ where $\tilde{g}$ is an unknown function independent on $x$, and
\begin{equation}
-\left\langle \vec{I}\right\rangle =\vec{\nu}\partial_{t}n_{d}+\vec{\nabla}j_{d}-\vec{\nabla}_{\bot}\tilde{g}
\label{<I>}
\end{equation}
Since $\partial_{x}\tilde{g}=0$, we can put $\tilde{g}=0$ for all problems where plastic deformations are localized within a limited part of the sample length which we shall assume further on. With the help of these relations we can average $\vec{I}$ in Eq.(\ref{tau+I}) to arrive at
\begin{equation}
\partial_{x}\langle\omega_{t}\rangle=
\partial_{t}\langle\omega_{x}\rangle-2\pi\frac{dn_{d}}{dt}\, \mathrm{or} \,
\partial_{x}\left[\langle\omega_{t}\rangle+2\pi j_{d}\right] =
\partial_{t}\left[ \langle\omega_{x}\rangle-2\pi n_{d}\right]
\label{omega-av}
\end{equation}
\begin{equation}
\partial_{y}\langle\omega_{t}\rangle
=\partial_{t}\langle\omega_{y}\rangle-2\pi\partial_{y}j_{d}\, , \,
\partial_{z}\langle\omega_{t}\rangle =\partial_{t}\langle\omega_{z}\rangle-2\pi\partial_{z}j_{d}  \notag
\end{equation}

The most important consequence of relations (\ref{<I>},\ref{omega-av}) is that we can introduce a uniquely defined function $\chi$ (which we shall
normalize as the phase measured in units of $\pi$) which derivatives are given as
\begin{equation}
\partial_{t}\chi=\langle\omega_{t}\rangle/\pi+2j_{d}\,~,\;\partial_{x}\chi=
\langle\omega_{x}\rangle/\pi-2n_{d}\,,~\partial_{y}\chi=
\langle\omega_{y}\rangle/\pi~,\;\partial_{z}\chi=\langle\omega_{z}\rangle /\pi
\label{chi}
\end{equation}
Discovering the uniquely defined average phase $\chi$ is equivalent to finding the first integral to four equations for $\omega_{j}$, $j=x,y,z,t$.

\section{Dynamics and averaging.}
\subsection{Local equilibrium.}

We shall employ specific units such that the electric potential $\Phi$ incorporates the electron charge $e$ and all energies are multiplied by the
factor $N_{F}$ - the density of states (per chain) at the Fermi level, so
that the dimension of energy becomes the inverse length. We shall employ the local energy functional appropriate to CDWs at presence of normal carriers \cite{SB+NK:19}:

\begin{align}
W\{\vec{\omega},\Phi,n\} & =\int dV\{\frac{1}{2\pi^{2}} \left[\omega_{x}^{2}+\alpha_{y}\omega_{y}^{2}+\alpha_{z}\omega_{z}^{2}\right] +{\frac{\omega_{x}}{\pi}}\Phi-\frac{1}{2}r_{0}^{2}(\nabla\Phi)^{2}  \label{W}
\\
& +\Phi n_{e}+\left(\Phi+\omega_{x}/\pi\right) n_{i}+\mathcal{F}_{c}(n_{i}^{e},n_{i}^{h},n_{e}^{e},n_{e}^{h})\}  \notag
\end{align}

Here $s$ is the area per one chain, $\varepsilon_{hst}$ is the dielectric susceptibility of the host crystal, $v_{F}$ is the Fermi velocity and and $r_{0}$ is the screening
length of the parent metal:  $r_{0}^{-2}={4\pi e^{2}N_{F}}/(\varepsilon_{hst}s)$.
$\mathcal{F}_{n}(n_{i}^{e},n_{i}^{h},n_{e}^{e},n_{e}^{h})$ is the free energy of normal carriers (electrons and holes) at their local equilibrium,
$n_{i,e}=n_{i,e}^{e}-n_{i,e}^{h}$ are concentrations of charges of intrinsic and extrinsic free carriers respectively.
There are two limiting types of free carriers: intrinsic ones which are common to all CDW materials at a finite temperature - their spectrum is formed by the CDW gap and moves as the Fermi level shifts with a breathing $n_{c}=\omega_{x}/\pi$ due to elastic deformations; extrinsic carriers may be also present belonging to other bands well decoupled from the vicinity of the CDW gap.
In addition to $\Phi$, we introduce two other potentials:
\begin{equation}
U=\pi\frac{\partial W}{\partial\omega_{x}}=\Phi+\frac{\omega_{x}}{\pi}+n_{i}~,~
V=\frac{\partial W}{\partial n_{i}}=\Phi+\frac{\omega_{x}}{\pi}
\label{U-V}
\end{equation}
The potential $U$ is just the longitudinal stress $\sigma_{x}$, $2U$ is the energy paid to distort the CDW elastically by one period, hence $U$ refers
to one electron in the ground state. $V$ is the potential experienced by intrinsic free carriers; it differs from the electric potential $\Phi$ experienced by extrinsic carriers due to elastic CDW deformations.

The CDW dynamics, which is commonly overdamped, is governed by the local balance of forces
\begin{equation}
\vec{\nabla}\vec{\sigma}=\mathcal{F}_{hst}~,~
\vec{\sigma}=\pi\partial W/\partial\vec{\omega}
\label{str1}
\end{equation}
where $\vec{\sigma}$ is the CDW internal stress conjugated to the strain $\vec{\omega}/\pi$ and $\mathcal{F}_{hst}(\omega_{t})$ is a total force exerted by the host crystal. For the energy functional (\ref{W})
\begin{equation}
\sigma_{x}=\Phi+\omega_{x}/\pi+n_{i}~,~\sigma_{y}=\alpha_{y}\omega_{y}/\pi,~
\sigma_{z}=\alpha_{z}\omega_{z}/\pi  \notag
\end{equation}
Suggesting a linear dependence $\mathcal{F}_{hst}(\omega_{t})=F_{pin}+\gamma\omega_{t}/\pi$, where $F_{pin}$ is a remnant pinning force and
$\gamma$ is the CDW damping parameter, we can write Eq.(\ref{str1}) as

\begin{equation}
(\vec{\nu}\vec{\nabla})\Phi+(\hat{\nabla}\vec{\omega})/\pi+(\vec{\nu}\vec{\nabla})n_{i}
=F_{pin}+\gamma\omega_{t}/\pi~,~\gamma\propto/\sigma_{c}
\label{str}
\end{equation}
where
$\hat{\nabla}=\left(\partial_{x},\alpha_{y}\partial_{y},\alpha_{z}\partial_{z}\right)$
and $\sigma_{c}$ is the collective conductivity. The Poisson equation $\delta W/\delta\Phi =0$ reads
\begin{equation}
r_{0}^{2}\Delta{\Phi}+n_{n}+\omega_{x}/\pi=0~,~n_{n}=n_{i}+n_{e}
 \label{Q}
\end{equation}

\subsection{Equations for the averaged quantities.}
Eqs.(\ref{str}) and (\ref{Q}) with the help of (\ref{chi}) allow us to obtain equations for the average phase $\chi$ and the potential $\Phi$
\begin{equation}
\left( \hat{\Delta}-\gamma\partial_{t}\right) \chi=F_{pin}+E-2\gamma j_{d}-\partial_{x}(n_{i}+2n_{d})
\label{Phi-chi}
\end{equation}

\begin{equation}
r_{0}^{2}\Delta{\Phi}+\partial_{x}\chi+2n_{d}+n_{n}=0~,~E=-\partial_{x}\Phi
\label{poisson}
\end{equation}
where
$\hat{\Delta}=\vec{\nabla}\hat{\nabla}=\partial_{x}^{2}+{\Delta}_{\bot} $,
${\Delta}_{\bot}=\alpha_{y}\partial_{y}^{2}+\alpha_{z}\partial_{z}^{2}$.
Eq. (\ref{Phi-chi}) tells that the phase $\chi$ is driven, in addition to standard forces $F_{pin}+E$, also by the current of defects and by the longitudinal gradient of the total number of particles. The canonical form of Eq. (\ref{Phi-chi})

\begin{equation}
-\gamma\partial_{t}\chi+\nabla\vec{\sigma}=F_{pin}-2\gamma j_{d}
\label{canonic}
\end{equation}
shows that the local equilibrium ($\chi=cnst$, $\vec{\sigma}=0$) is perturbed only by the current of defects representing the glide of dislocations.

It follows from Eq. (\ref{poisson}) that total densities of the charge $n_{tot}$ and hence of the current $\vec{j}_{tot}$ are given as
\begin{equation*}
n_{tot}=\omega_{x}/\pi+n_{n}=\partial_{x}\chi+2n_{d}+n_{n}~\mathrm{then}~
\vec{j}_{tot}=(-\partial_{t}\chi+2j_{d})\vec{\nu}+\vec{j}_{n}
\end{equation*}

The density and the current of defects contribute in the frame of the average phase $\chi$ while they were doomed with respect to local deformation as it is seen from Eq.(\ref{Q}), thus resolving the paradox outlined in Sec.\ref{CDW-Xl}. The conservation law for the total number of particles reads
\begin{equation}
\frac{dn_{d}}{dt}=-2\frac{dn_{n}}{dt}=R~,~~\frac{dn_{n}}{dt}=\frac{\partial n_{n}}{\partial t}+\vec{\nabla}\vec{j}_{n}
\label{n-tot}
\end{equation}
where the balance between the normal and the condensed carriers is maintained by the microscopically defined conversion rate $R$ which is controlled by a mismatch of chemical potentials $\mu_{d}$ and $\mu_{e,i}$
\begin{equation}
\mu_{a}=\frac{\delta W}{\delta n_{a}}=V_{a}+\zeta_{a}~,~
\zeta_{a}=\frac{\partial W}{\partial n_{a}}~,~
V_{e}=\Phi\,~,~V_{i}=V=\Phi+\partial_{x}\chi+2n_{d}
\label{j-mu}
\end{equation}
Here $V_{a}$ are appropriate potentials and $\zeta_{a}(n_{a})$ are the local chemical potentials depending on concentrations. In the diffusion approximation, the currents of normal carriers are given as
$\vec{j}_{a}=-\hat{G}_{a}\nabla\mu_{a}$, $a=e,i$ where $\hat{G}_{a}$ are the conductivity tensors.

At first sight, the force acting upon a nucleus defect, the $2\pi$ soliton, might also follow the gradients of their potential energy
\begin{equation*}
V_{d}=2U~,~U=\Phi+\partial_{x}\chi+n_{i}+2n_{d}
\end{equation*}
But it would be a mistake to use $-2\partial_{x}U$ as the driving force like for other potentials in (\ref{j-mu}).
The force acting upon the gliding loop is not given by the energy gradient as supposed by analogy with magnetic forces upon currents:
\begin{align}
\mathcal{F}_{x}/2 & =\oint[\vec{\nu}\times~\vec{\sigma}]d\vec{l}=
\int\{\vec{\nu}(\nabla\vec{\sigma})-(\nabla\vec{\nu})\vec{\sigma}\}d\vec {s}
\label{F-glide} \\
& \Rightarrow\int ds(\nabla_{\perp}\vec{\sigma})=\int ds(\mathcal{F}_{hst}-\partial_{x}U)
\notag
\end{align}
where expressions in the second line correspond to a plane loop.
Because of the interaction (including the pinning and the friction) with the host lattice, $\vec{\nabla}\vec{\sigma}=F_{hst}\neq0$ (\ref{canonic}), so the force is not given by the expected \cite{LLv7} for conventional crystals gradient of the potential $2U$: the force driving the glide of defects comes only from share strains - the transverse gradients
\begin{equation}
-\partial_{x}V_{d}\Rightarrow2\hat{\Delta}_{\bot}\chi\neq-\partial_{x}V_{d}~,~
j_{d}=-2G_{d}(\hat{\Delta}_{\bot}\chi+\partial_{x}\zeta_{d})
\label{j_d}
\end{equation}

Recall for completeness the counterpart of (\ref{F-glide}) - the force $F_{\bot}\bot\vec{b}$ (written here for a plane loop):
\begin{equation*}
F_{\bot}=2\oint\sigma_{x}\vec{\nu}\times d\vec{l}=2\int
ds\nabla_{\perp}\sigma_{x}
\end{equation*}
which reminds that $2U$ is the energy paid for expansion of a loop per one added chain. Being interpreted as a force, $F_{\bot}$ would promote the climb if the dislocation line expansion is not forbidden by the charge conservation law.
Recall the common prescription \cite{LLv7} to project this motion out. Here, we must allow for the climb while considering it not as a motion under certain forces but as a growth by adhesion of normal carriers which is controlled by the conversion rate function $R$ as in (\ref{n-tot}).

Let $n_{\pm}$, with $n_{+}-n_{-}=n_{d}$, $n_{+}+n_{-}=n_{d,tot}$ are the partial concentrations of defects with two signs of vorticity,
$n_{\infty}$ is their total concentration $n_{d,tot}$ in the bulk equilibrium where the chemical potential $\zeta_{d}=0$; the forces acting upon defects with
vorticities $\pm$ are $\pm F=\mp\hat{\Delta}_{\perp}\chi$. The current of defects is, in the diffusion approximation,
\begin{equation}
j_{d}=b_{d}F(n_{+}+n_{-})-D_{d}\partial_{x}(n_{+}-n_{-})=b_{d}n_{d,tot}(\hat{\Delta}_{_{\perp}}\chi-\partial_{x}\zeta_{d})
\label{j+n,+-}
\end{equation}
\begin{equation}
n_{\pm}=n_{\infty}/2\exp\left(\pm{\zeta_{d}}/{T}\right)
\end{equation}
where $b_{d}$ and $D_{d}=b_{d}T$ are the mobility and the diffusion coefficient of defects. Substituting this expression to Eq.(\ref{Phi-chi}) we get
\begin{equation}
\left( \hat{\Delta}-\gamma\partial_{t}\right) \chi+2\gamma b_{d}n_{d,tot}\hat{\Delta}_{\bot}\chi+
2(\gamma D_{d}+1)\partial_{x}n_{d}=F_{pin}+E
\end{equation}
In a sense, the allowance for defects' motion contributes additively to the transverse rigidity $\propto\hat{\Delta}_{\bot}\chi$ of the phase and to the driving force from the gradient of the defects' concentration.

\subsection{The direct method of averaging and an extended dislocations.}
It is instructive to demonstrate an alternative method of deriving equations for average fields based on explicit equations of motion.
Differentiating Eq. (\ref{str}) over $x$ and using relations (\ref{tau+I}) we get the equation for $\omega_{x}$
\begin{equation}
\left( \hat{\Delta}-\gamma\partial_{t}\right) \omega_{x}/\pi+2[\hat{\nabla\times\vec{\nu}}]\vec{\tau}-
	2\gamma\partial_{t}(\vec{\nu}\vec{P}){-}(\vec{\nu}\vec{\nabla})F_{pin}+
	{(\vec{\nu}\vec{\nabla})^{2}}n_{i}+({\vec{\nu}\vec{\nabla})^{2}\Phi}=0
	\label{omega-Phi}
\end{equation}
In average
\begin{equation}
\left\langle\left(\lbrack\hat{\nabla}\times\vec{\nu}]\vec{\tau}\right)\right\rangle =-\hat{\Delta}_{\bot}n_{d}
\label{R-n}
\end{equation}
then
\begin{equation}
\left( \hat{\Delta}-\gamma\partial_{t}\right) \omega_{x}/\pi-2\hat{\Delta}_{\bot}n_{d}-2\gamma\frac{dn_{d}}{dt} - (\vec{\nu}\vec{\nabla})F_{pin}+{(\vec{\nu}\vec{\nabla})^{2}}n_{i}+({\vec{\nu}\vec{\nabla})^{2}\Phi}=0
\end{equation}
and with the help of $\partial_{x}\chi+2n_{d\,}=\langle\omega_{x}\rangle/\pi$ we have
\begin{equation}
\hat{\Delta}\partial_{x}\chi+2\partial_{x}^{2}n_{d}-\gamma\partial_{t}\partial_{x}\chi-2\gamma\partial_{x}j_{d}{-}\partial_{x}F_{pin}+
\partial_{x}{^{2}}n_{i}+\partial_{x}{^{2}\Phi}=0
\end{equation}
The LHS is the full derivative over $x$; then, up to an arbitrary $x$-independent function, the first integral yields again Eq. (\ref{Phi-chi}).
The presence of noncompensated dislocation lines $\vec{\tau}_{D}$ on top of the ensemble of loops or pairs can be taken into account by generalizing Eq.
(\ref{<tau>}) as
$[\vec{\nu}\vec{\nabla}]n_{d}\Rightarrow\lbrack\vec{\nu}\vec{\nabla}]n_{d}-\vec{\tau}_{D}$
and substituting that to Eq. (\ref{Phi-chi}) after application of the operator $[\vec{\nu}\vec{\nabla}]$:
\begin{equation}
[\vec{\nu}\vec{\nabla}][\left( \hat{\Delta}-\gamma\partial_{t}\right) \chi+2\gamma j_{d}+\partial_{x}(n_{i}+2n_{d})-F_{pin}-E]
=2\partial_{x}\vec{\tau}_{D}
\label{Disl}
\end{equation}

\section{Limiting cases and applications.}

\subsection{Electroneutrality approximation and the 1D regime.}
Usually $r_{0}\sim nm$ is the smallest characteristic length in the system,
then in lowest orders of $r_{0}$ we can write a separate equation for $\chi$:
\begin{equation}
[r_{0}^{2}\Delta_{\bot}\left( \hat{\Delta}_{\bot}-\gamma\partial_{t}\right) -\partial_{x}^{2}]\chi
=\partial_{x}(n_{n}+2n_{d})+2\gamma j_{d}
\label{L0-chi}
\end{equation}
while $\Phi$ can be restored via the Eq. (\ref{Phi-chi}). Integrating over the sample cross-section, the equations for integrated $\chi$ and $n$ are reduced to simple connections
\begin{equation}
\partial_{x}\bar{\chi}+\bar{n}_{n}+2\bar{n}_{d}=0~,~ -\partial_{t}\bar{\chi}+\bar{j}_{n}+2\bar{j}_{d}=J(t)
\label{chi-1d}
\end{equation}
meaning the electroneutrality and the total current conservation. They just indicate that within the electroneutrality approximation the charge integrated over any cross-section is equal zero and the total current cannot depend on $x$: $J=J(t)$. The equation connecting integrated $\bar{\chi}$ and $\bar{\Phi }$ reads
\begin{equation}
E+\partial_{x}\bar{n}_{e}+F_{pin}=-\gamma\partial_{t}\bar{\chi}+2\gamma\bar {j}_{d}=-\gamma\bar{j}_{n}+\gamma J(t)
\label{pot-1d}
\end{equation}
These relations tell, not quite predictably, that in average over the crossection the electric field opposes the pinning, the gradient of
(extrinsic only) normal carriers and the friction from abnormal (of overall sliding and of defects) currents.
Relations between integrated potentials $\Phi,U,V$ (\ref{U-V}) acquire the simple form:
$\bar{U}=\bar{\Phi}-\bar{n}_{e}$, $\bar{V}=\bar{\Phi}-\bar{n}_{n}$.

For a gapful CDW where $n_{e}=0$ and also at low $T\ll\Delta$ when $n_{i}$ can be neglected, the only carriers are the phase defects.
Unlike for normal carriers, the specific driving force (\ref{j+n,+-}) of defects vanishes after the integration in view of the zero-stress side boundary conditions $\nabla_{\perp}\chi|_{sides}=0$.
Hence the integrated current of defects is driven only by the diffusion: $\bar{j}_{d}\approx-D_{d}\partial_{x}\bar{n}_{d}$, $(\partial_{t}-D_{d}\partial_{x}^{2})\bar{n}_{d}=0$.
Then with Eq. (\ref{chi-1d})
\begin{equation}
\partial_{x}\bar{\chi}+2\bar{n}_{d}=0~,~\partial_{t}\bar{\chi}+2D_{d}\partial_{x}n_{d}=-J(t)~,~(D_{d}\partial_{x}^{2}-\partial_{t})\bar{\chi}=J(t)\\
E+F_{pin}=\gamma J(t)
\end{equation}
These seemingly simple relations show actually an unexpected result for the elastic response $\partial_{x}^{2}\bar{\chi}$ which is a well measurable quantity, once having attracted a big experimental attention (see \cite{Lemay:98,Rideau:01} and refs. therein).
This response with respect to the current $J$ is given curiously by the diffusion coefficients of defects.
The response to the electric field is not given by the bare elastic part of Eq. (\ref{Phi-chi}) ($E$ versus $\partial_{x}^{2}\chi$) but versus $\gamma D_{d}$.
That is the conventionally thermodynamic elasticity becomes governed rather by the product of kinetic coefficients: collective mode friction and the diffusion of its nonlinear excitations.
We can take into account also the normal carriers in their dynamic equilibrium with defects. Linearizing relations (\ref{n-tot},\ref{j-mu}) as
\begin{equation}
n_{a}\approx\zeta_{a}\rho_{a}~,~R\approx(\mu_{d}-2\mu_{n})/\tau
\end{equation}
where $\rho_{a}=cnst$ and $\tau$ is a mean time for recombination of two electrons or holes into a growing dislocation loop, we can arrive at
equations \cite{BK:99} which have been already exploited \cite{Rideau:01} in explaining experiments as shown in Fig.2.
In applications we should keep in mind that measured quantities are not given just by derivatives of the phase $\chi$ but by the local derivatives: $\omega_{x}$ for the CDW wave vector shift $q$ plotted in Fig.2 and $\omega_{t}$ for the CDW velocity as it was measured by the coherent generation of the "narrow band noise".

\subsection{ An isolated dislocation.}
Consider now a static dislocation in presence of the gas of point defects - the solitons. Let the dislocation line is directed in $z$ being centered at $(X_{D},Y_{D})=(0,0)$, then Eq.(\ref{Disl}) in the approximation of Eq. (\ref{L0-chi}) takes the form
\begin{equation}
\lbrack r_{0}^{2}\alpha\partial_{y}^{4}-\partial_{x}^{2}]\chi-
\partial_{x}{(}n_{n}+2n_{d})=\partial_{x}\delta(x)\mathrm{Sgn}(y)~,~\alpha=\alpha_{y}
\label{disl-EN}
\end{equation}
The electric potential can be restored from a solution of Eq.(\ref{disl-EN}) via the equation
\begin{equation}
-\partial_{x}\Phi+\partial_{x}n_{e}(\mu_{e}-\Phi)=\alpha\partial_{y}^{2}\chi
\label{D:chi,nd,E}
\end{equation}
From now on, the singular source terms like in the RHS of Eq.(\ref{disl-EN}) will be omitted assuming appropriate boundary conditions as it will indicated below.

In the following we shall concentrate on most interesting effects of solitons, neglecting contributions from normal carriers which is valid for a
typical gapful CDW with $n_{e}=0$ at low $T\ll\Delta$ when $n_{i}\propto\exp(-\Delta/T)$ also freezes out.
Henceforth we shall omit the index "$_{d}$": $n_{d}\rightarrow n$, $\zeta_{d}\rightarrow\zeta$.
Under the equilibrium condition $j_{d}=0$ Eq.(\ref{j+n,+-}) yields
\begin{equation}
n=n_{\infty}\sinh\frac{\zeta}{T}~,~\partial_{x}\zeta=\Phi=
\alpha\partial_{y}^{2}\chi
\label{zeta-pot}
\end{equation}
With the help of (\ref{zeta-pot}) the degree of derivatives in Eq.(\ref{disl-EN}) can be reduced as
\begin{equation}
2n_{\infty}\left(l_{scr}^{2}\partial_{y}^{2}\frac{\zeta}{T}-\sinh\frac{\zeta}{T}\right) -\partial_{x}\chi=0
~,~ l_{scr}=r_{0}\left( \frac{T}{2n_{\infty}}\right) ^{1/2}
\label{l_scr}
\end{equation}
where $l_{scr}$ is the Debye screening length provided by defects.

At the most distant region where $|\zeta|\ll T$ we can linearize the last equation as $\sinh z\rightarrow z$.
Also at $|y|\gg l_{scr}$ we can neglect the highest derivative term $l_{scr}^{2}\partial_{y}^{2}$.
Then we arrive at a kind of a normal elastic theory governed by a stretched Laplacian
\begin{equation}
\partial_{x}^{2}+\beta^{2}\partial_{y}^{2}~,~
\beta=\sqrt{\alpha}(r_{0}/l_{scr})=\sqrt{2\alpha n_{\infty}/T}\ll1
\end{equation}
where the effective elastic anisotropy $\beta^{2}$ is determined by the concentration of defects and $\beta$ vanishes when the carriers freeze out.
The phase distribution is a vortex $\chi=\arctan(\beta x/y)$ which strong elongation in the $x$ direction progresses with decreasing of $n_{\infty}$.
In this regime
\begin{equation}
n_{d}\approx n_{\infty}\frac{\zeta}{T}=\beta^{2}\int\partial_{y}^{2}\chi dx=-\partial_{x}\chi=\frac{y\beta/\pi}{y^{2}+\beta^{2}x^{2}}
\end{equation}

Within the screening length $|y|<l_{scr}$ the term with high $y-$ derivatives in (\ref{l_scr}) becomes important and we enter the regime of the anomalous elasticity dominated by unscreened Coulomb interactions \cite{SB+SM-disl:91,SB+SM-sol:91}.
Now the solution of (\ref{l_scr}) together with (\ref{zeta-pot}), or equivalently of Eq.(\ref{disl-EN}) neglecting $n_{n}$ and $n_{d}$, becomes
\begin{eqnarray}
\partial_{y}\chi=\sqrt{\frac{1}{4\pi d|x|}}\exp\left(-\frac{y^{2}}{4d|x|}\right)\mathrm{Sgn}(x)~,~
d=\sqrt{\alpha}r_{0}  \notag \\
\zeta=\frac{\sqrt{\alpha}}{2r_{0}}\chi(x,y)\mathrm{Sgn}(x)=
\frac{\alpha^{1/2}}{4r_{0}}\mathrm{erf}\left(\frac{y}{2\sqrt{d|x|}}\right)\mathrm{Sgn}(y)
\label{dchidy}
\end{eqnarray}
The expression (\ref{dchidy}) satisfies the condition $\chi(x,\infty)-\chi(x,-\infty)=\mathrm{Sgn}(x)$ thus defining the $2\pi$ circulation of the phase $\pi\chi$ with an essential peculiarity that the vorticity is concentrated in two parabolic wings $|y|<\sqrt{d|x|}$ spread from the dislocation.
By transversing the wings, the potentials $\zeta\approx\Phi$ acquire a magnitude (in dimensional units of the energy) $\Phi_{0}=\alpha^{1/2}/(4r_{0}N_{F})\sim\alpha^{1/2}\hbar\omega_{p}$,  $\Phi_{0}/T_{c}\sim\omega_{p}/\Delta$
which scales as a plasma frequency $\omega_{p}$ of the parent metal which in typical CDWs is the biggest energy $\hbar\omega_{p}>\Delta>T_{c}>T$.
Hence the linearisation of the Eq.(\ref{l_scr}) leading to the analytical results (\ref{dchidy}) is valid only
close to the axes of the wings at $|y|\ll\sqrt{d|x|}$ where
$\chi\approx y/\sqrt{4\pi d|x|}$, $\zeta\approx y/(2r_{0})\sqrt{\alpha/(\pi |x|)}$
i.e. in the regime where the transverse electric field $E_{y}$ is a constant in the transverse direction $y$ with a slowly decaying magnitude
$E_{y}\sim(4\pi r_{0}|x|)^{-1}$ along the chains.

The above analytic solutions of linearized equations are valid only at large distances from the dislocation line.
The nonlinear regime covers the important region $(n_{\infty}x)^{2}<|y|n_{\infty}/\beta<1$ which is adjacent, while still wide, to the dislocation core.
Here only a numerical solution is possible which results are present below. We introduce dimensionless  variables
\begin{equation}
z=\zeta/T~,~\xi=x2n_{\infty}~,~\eta=y(2Tn_{\infty})^{1/2}=y/l_{scr}=y2n_{\infty}/c
\end{equation}
to simplify the system of equations as
\begin{equation}
\partial_{\xi}\chi=c^{2}\partial_{\eta}^{2}Z-\sinh(z)
~,~\partial_{\xi}Z=\partial_{\eta}^{2}\chi
\label{2z-1chi}
\end{equation}

The singular source terms in the RHSs of the original equations will be emulated by discontinuous boundary conditions compatible with the
traditional vortex form $\pi\chi=\arg(\xi+i\eta)$ valid at large distances:
$\chi(0,y)=1/2\mathrm{Sgn}(y)$, $\partial_{x}\zeta(0,y)=0$, $\chi(x,0)=\Theta(-x)$, $\zeta(x,0)=0$.

Results of a numerical solution of the system of partial differential equations (\ref{2z-1chi}) are illustrated in Fig.5  (for $c =0.1$).
Here the dislocation is centered at $(\xi =0,\eta =0)$ and only the upper half-plane $\eta >0$ is shown;
the functions $Z(\xi,\eta),\chi (\xi ,\eta)$ are continued anti-symmetrically in  $\eta $ so that $Z(\xi ,\eta )$ is continuous through $\eta =0$ while
$\chi (\xi ,\eta )$ experiences a jump over the semi-axes $\eta =0,\xi <0$.
Vectors $\{\cos (\pi\chi ),\sin (\pi \chi)\}$ and streamlines of $[\vec{\nu}\nabla ]\chi$ characterize the phase $\pi \chi $.
The color indicates the chemical potential $\zeta =ZT$ where $Z$ changes from $\approx 0$ at large distances to a maximal value $Z\approx 2.5$ approaching the dislocation and then drops to zero in the core.
Thanks to the nonlinearity taken into account, there is no more nonphysical divergence of $Z$ which was inherent to the linearized equations.
Still the enhancement up to $Z\approx 2.5$ is appreciable corresponding to increasing of solitons' concentration near the dislocation by a factor $n(core)/n_{\infty }\approx6$.
The conventional rotation of the phase following the coordinate angle at large distances becomes near the core a nearly vertical drop indicating the high $x$- gradient in accordance with the rapidly growing $Z$.

\section{Conclusions.}

As a kind of a crystal of electronic pairs, the incommensurate CDW possess a translational symmetry (appearing at the microscopic scale as a chiral one)
which gives rise to the sliding mode and to topological defects such as electronic vortices - the dislocations, their rings or pairs.
In a quasi one-dimensional system, which is a typical playground for CDWs, the minimal ring becomes the $2\pi$ on-chain soliton taking the role of major charge
carriers.
With a dominant role of Coulomb interactions, the screening facilities of solitons are ultimately important for the structure and energetics of macroscopic dislocations. Moreover, the solitons feed the transverse proliferation, the climb, of dislocations giving rise to phase slip events resulting in the overall sliding.
These connections require for a hydrodynamic theory embracing the electric field and elastic distortions, ensembles of normal carriers and dislocation loops.
Particular care must be taken to introduce a phase allowable for the ensemble averaging, to single out the charges and currents of solitons with respect to the ones from
collective contributions and to understand forces driving the current of solitons.
The presented theory provides a general frame to address these issues and describes a basic scenario of a gas of solitons at presence of a macroscopic dislocation.
For further applications, the theory still needs to take into account the distribution, and its evolution, of loops' dimensions to describe their aggregation towards macroscopic objects.

\newpage
\textbf{Figure Captions}
\\  \\
Fig.1. STM image of CDWs at chains of $NbSe_{3}$ compounds \cite{Brun-soliton}. An extra CDW periodicity is observed (inside the ellipse) at the location of which the soliton is trapped.\\
\\
Fig.2. Profile of the CDW phase gradient $\partial_{x}\varphi$
deformed in the course conversion among normal and collective currents near
junctions ($x\approx\pm2$) and at an obstacle ($x=0$). The two families of
data (shown as circles and stars) correspond to opposite directions of the
applied current. The dashed line is the theory fit.\\
\\
Fig.3. The surface $S$ of phase discontinuities $\vec{P}$ based upon the
dislocation loop $\vec{\tau}$ with circulations of the phase
gradient $\vec{\omega}$.\\
\\
Fig.4.	Left: Propagation of the dislocation loop in the course of its glide producing the current $j_{d}$ in chains' direction. The arrows show the vectors $\vec{P}$ distributed over the surface of the phase discontinuities.\\
Right: Expansion of the dislocation loop in the course of its climb.
Thin arrows show the climb direction. The surface vector $\vec{P}$ is perpendicular to the plane. \\
\\
Fig.5. \\
Distributions around a dislocation centered at (0,0) (only the
upper half-plane is shown). $\xi,\eta$ are dimensionless
rescaled coordinates $x,y$. Vectors and streamlines characterize the phase $\pi\chi$. The color indicates the chemical potential $\zeta=ZT$. $Z$ changes from $\approx 0$ at large
distances (green color) to a maximal value $Z\approx 2.5$ near the origin (red
color) and then drops to zero (blue color).

\newpage
Figure1.
\begin{figure}[tbh]
	\centering
	\includegraphics{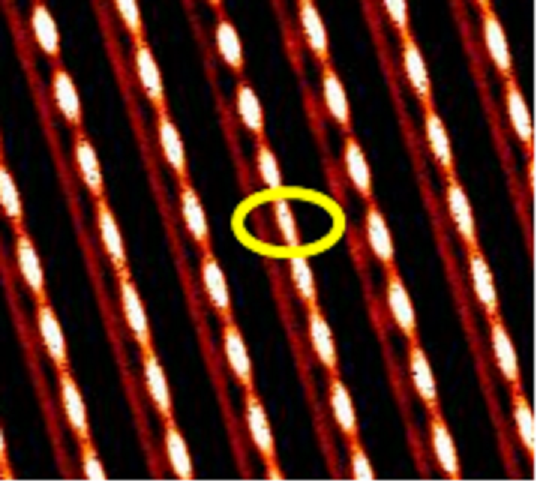}
	\label{fig:STM}
\end{figure}

\newpage
Figure 2.
\begin{figure}[tbh]
	\centering
	\includegraphics{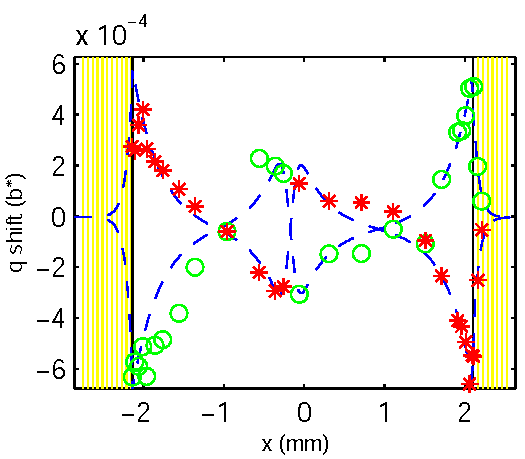}
\end{figure}
\newpage

\newpage
Figure 3.
\begin{figure}[tbh]
	\centering
	\includegraphics{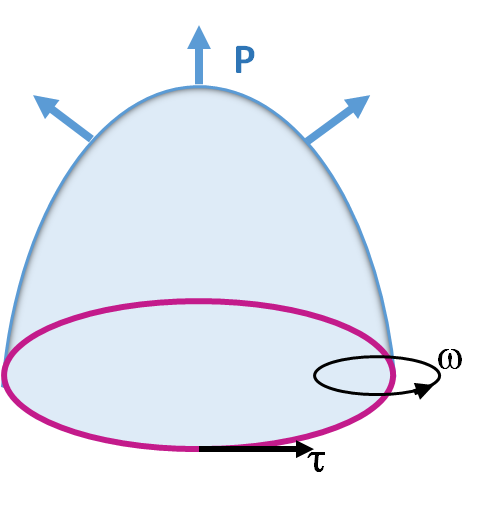}
\end{figure}

\newpage
Figure 4.\\
\\
\begin{figure}[tbh]
	\centering
	\includegraphics{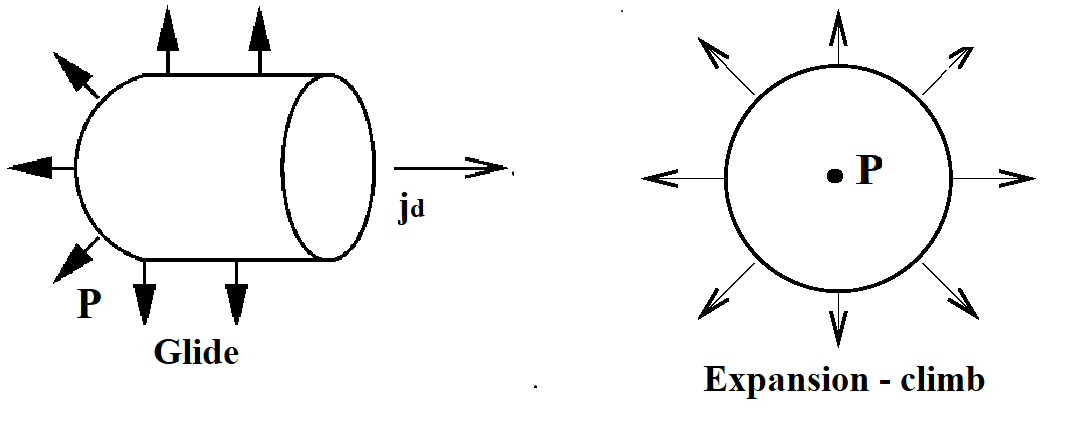}
\end{figure}

\newpage

Figure 5.
\begin{figure}[tbh!]
	\centering
	{\includegraphics[width=10cm]{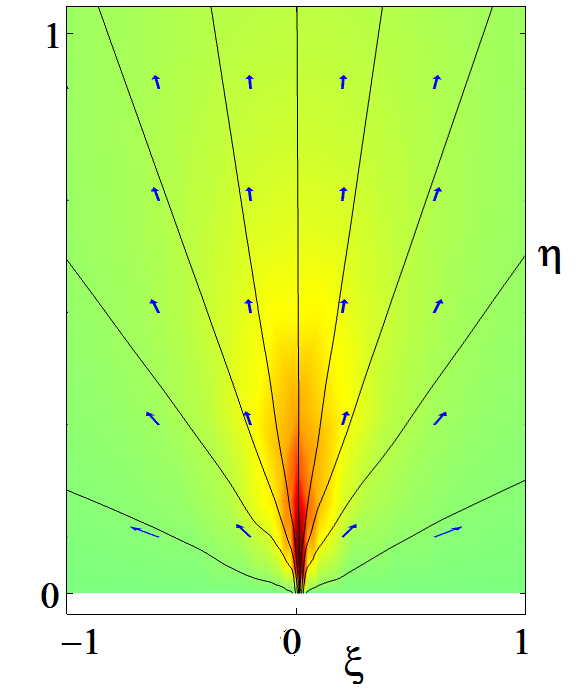}}
\end{figure}

\end{document}